\documentclass[prl,showpacs,twocolumn,floats,10pt,aps,citeautoscript,longbibliography,superscriptaddress,nobalancelastpage]{revtex4-2}
\usepackage{dcolumn}
\usepackage{anyfontsize} 
\usepackage{microtype} 
\usepackage{graphicx} 
\usepackage{xspace} 

\usepackage[usenames,dvipsnames]{xcolor} 

\usepackage{xparse} 

\usepackage{algorithm}
\usepackage{algpseudocode}

\usepackage[normalem]{ulem} 
\definecolor{darkgreen}{rgb}{0,0.5,0}
\definecolor{purple}{rgb}{0.6,0,0.5}
\definecolor{orange}{rgb}{1,0.5,0}
\definecolor{darkred}{rgb}{.7,0,0}
\definecolor{darkblue}{rgb}{0,0,.6}
\definecolor{grey}{rgb}{.6,.6,.6}
\definecolor{dimgreen}{rgb}{0.2,0.7,0.2}

\newcommand{\jvdomit}[1]{}

\usepackage{amsmath} 
\usepackage{amssymb}
\usepackage{mathrsfs} 
\usepackage{bbm}  
\renewcommand{\vec}[1]{{\mathbf{#1}}} 

\allowdisplaybreaks 
\usepackage{enumitem} 

\newcommand{\Eq}[1]{Eq.~\eqref{#1}}

\newcommand{\Fig}[1]{Fig.~\ref{#1}}
\newcommand{\Figs}[1]{Figs.~\ref{#1}}

\usepackage{scalefnt} 

\def\Cc{\mathcal{C}}

\def\bk{\mathbf{k}}
\def\bx{\mathbf{x}}
\def\bu{\mathbf{u}}

\def\bsigma{{\boldsymbol{\sigma}}}

\newcommand{\ellplusone}{{\ell  \hspace{-0.2mm}+\hspace{-0.2mm} 1}}

\newcommand{\eLL}{{\mbox{\small$\mathscr{L}$}}}
\newcommand{\seLL}{{\scriptscriptstyle \! \mathscr{L}}}
\newcommand{\scripteLL}{{\scriptstyle \mathscr{L}}}
\newcommand{\sseLL}{{{\mbox{\tiny$\!\mathscr{L}$}}}}

\newcommand{\eDD}{{\mbox{\small$\mathscr{D}$}}}
\newcommand{\crit}{{\mathrm c}}

\def\At{{\tilde{A}}}

\def\Dmax{D_{\mathrm{max}}}

\newcommand*{\ndots}{\kern-0.075em.\kern-0.05em.\kern-0.05em.}  
\newcommand*{\nidots}{.\kern-0.05em.\kern-0.05em.} 
\newcommand*{\ncdots}{\kern-0.15em\cdot\kern-0.2em\cdot\kern-0.2em\cdot\kern-0.15em}  

\NewDocumentCommand{\doubleI}{O{}}{\mathbbm{1}_{#1}}
\NewDocumentCommand{\doubleIb}{O{}}{{\overline{\mathbbm{1}}_{#1}}}
\NewDocumentCommand{\doubleIk}{O{}}{\mathbbm{1}^\ks_{\! #1}}
\NewDocumentCommand{\doubleId}{O{}}{\mathbbm{1}^\ds_{\! #1}}
\NewDocumentCommand{\doubleIp}{O{}}{\mathbbm{1}^\ps_{\! #1}}
\NewDocumentCommand{\doubleV}{O{}}{\mathbbm{V}_{\! #1}}
\NewDocumentCommand{\doubleVk}{O{}}{\mathbbm{V}^\ks_{\! #1}}
\NewDocumentCommand{\doubleVd}{O{}}{\mathbbm{V}^\ds_{\! #1}}
\NewDocumentCommand{\doubleVp}{O{}}{\mathbbm{V}^\ps_{\! #1}}
\NewDocumentCommand{\doublev}{o}{{\mathbbm{v}_{#1}}}
\NewDocumentCommand{\doubleVb}{o}{{\overline{\mathbbm{V}}_{\! #1}}}
\NewDocumentCommand{\doubleVt}{o}{{\widetilde{\mathbbm{V}}_{\! #1}}}
\NewDocumentCommand{\doubleVh}{o}{\widehat{{\mathbbm{V}}_{\! #1}}}
\NewDocumentCommand{\doubleW}{o}{\mathbbm{W}_{\! #1}}
\NewDocumentCommand{\doubleWk}{o}{\mathbbm{W}^\ks_{\! #1}}
\NewDocumentCommand{\doubleWd}{o}{\mathbbm{W}^\ds_{\! #1}}
\NewDocumentCommand{\doubleWb}{o}{{\overline{\mathbbm{W}}_{\! #1}}}
\NewDocumentCommand{\doubleWt}{o}{{\widetilde{\mathbbm{V}}_{\! #1}}}
\NewDocumentCommand{\doubleWh}{o}{{\widehat{\mathbbm{V}}_{\! #1}}}

\def\ds{{\scriptscriptstyle {\rm D}}}
\def\ks{{\scriptscriptstyle {\rm K}}}
\def\ps{{\scriptscriptstyle {\rm P}}}

\def\ps{{\scriptscriptstyle {\rm P}}}

\newcommand{\chern}{\Cc}
\newcommand{\abs}[1]{{\left\lvert#1\right\rvert}}
\newcommand{\norm}[1]{{\left\lVert#1\right\rVert}}
\newcommand{\order}{\mathcal{O}}
\newcommand{\deltam}{\delta_m}
\newcommand{\berry}{{F}}

\newcommand{\braket}[2]{\langle#1\vert#2\rangle}

\newcommand{\ket}[1]{\lvert#1\rangle}
\newcommand{\rank}{\operatorname{rank}}
\newcommand{\Tr}{\operatorname{Tr}}

\newcommand{\initsigma}{\bar{\sigma}}

\usepackage{xparse} 

\RequirePackage[
  hyperindex,colorlinks,bookmarksnumbered,
  plainpages=true,pdfstartview=FitH]{hyperref}
\hypersetup{linkcolor=blue,urlcolor=blue,citecolor=blue}
\usepackage{hyperref}
\usepackage[all]{hypcap}

\usepackage{orcidlink}
\usepackage{physrevsupplement}

\begin{document} 

\title{Quantics Tensor Cross Interpolation for High-Resolution, Parsimonious Representations of Multivariate Functions in Physics and Beyond}
\author{Marc K.\ Ritter\,\orcidlink{0000-0002-2960-5471}}
\affiliation{Arnold Sommerfeld Center for Theoretical Physics, 
Center for NanoScience,\looseness=-1\,  and 
Munich Center for \\ Quantum Science and Technology,\looseness=-2\, 
Ludwig-Maximilians-Universit\"at M\"unchen, 80333 Munich, Germany}
\author{Yuriel \surname{N\'u\~nez Fern\'andez}\,\orcidlink{0000-0002-0080-1903}}
\affiliation{Universit\'e Grenoble Alpes, CEA, Grenoble INP, IRIG, Pheliqs, F-38000 Grenoble, France}
\author{Markus Wallerberger}
\affiliation{Institute of Solid State Physics, TU Wien, 1040 Vienna, Austria}
\author{Jan von Delft\,\orcidlink{0000-0002-8655-0999}}
\affiliation{Arnold Sommerfeld Center for Theoretical Physics, 
Center for NanoScience,\looseness=-1\,  and 
Munich Center for \\ Quantum Science and Technology,\looseness=-2\, 
Ludwig-Maximilians-Universit\"at M\"unchen, 80333 Munich, Germany}
\author{Hiroshi Shinaoka\,\orcidlink{0000-0002-7058-8765}}
\affiliation{Department of Physics, Saitama University, Saitama 338-8570, Japan}
\author{Xavier Waintal\,\orcidlink{0000-0003-3816-8290}}
\affiliation{Universit\'e Grenoble Alpes, CEA, Grenoble INP, IRIG, Pheliqs, F-38000 Grenoble, France}

\begin{abstract}
\begin{center}
(Dated: September 11, 2023)
\end{center}

Multivariate functions of continuous variables arise in countless branches of science.
Numerical computations with such functions typically 
involve a compromise between two contrary desiderata: accurate resolution of the functional dependence, versus parsimonious memory usage. Recently, two promising strategies have emerged for satisfying both requirements: 
(i) The  \textit{quantics} representation,
which expresses functions as multi-index tensors, with each index representing one 
bit of a binary encoding of one of the variables; and (ii) \textit{tensor cross interpolation} (TCI), which, if applicable, yields parsimonious interpolations for multi-index tensors. Here, 
we present a strategy, \textit{quantics TCI} (QTCI), which combines the advantages of both schemes. We illustrate its potential with an application from condensed matter physics: the computation of Brillouin zone integrals. 
\\
\\
\noindent
DOI: \href{https://doi.org/10.48550/arXiv.2303.11819}{10.48550/arXiv.2303.11819}
\end{abstract}

\vspace{-10mm}

\maketitle
\textit{Introduction.---}%
Let $f$ be a multivariate function of $n$ continuous,
real variables
$u_i$ ($i = 1, \dots, n$): \vspace{-1mm}
\begin{align}
	f \! : \! U \subset \mathbbm{R}^n \to \mathbbm{C} , 
	\quad 
  \bu = (u_1,  \dots , u_n) \mapsto f(\bu) \, . 
\end{align}
Such functions arise in essentially all branches of science. In physics, e.g., they could stand for the fields  used in classical or quantum field theories, with $\bu = (\bx,t)$ or 
$\bu = (\bk, \omega)$ representing space-time or momentum-frequency variables
in $n = \eDD+1$ dimensions, respectively; or for $m$-point correlation functions of such fields, with  $\bu = (\bx_1, t_1, \dots, \bx_m, t_m)$ and $n = m(\eDD+1)$, etc.  

Often such functions have structure (peaks, wiggles, divergences, even discontinuities) on length/time scales or momentum/frequency scales differing by orders of magnitude. Then, their numerical treatment 
is challenging due to two contrary requirements:  
On the one hand, accurate resolution of small-scale structures requires a fine-grained discretization grid,
while large-scale structures require a large domain of definition $U$;
and on the other hand, memory usage should be parsimonious, hence a fine-grained grid cannot be used throughout $U$.
In practice, compromises are needed, sacrificing resolution and/or restricting $U$ to limit memory costs, or using nonuniform grid spacings to resolve some parts of $U$ more finely than others.

Very recently, in different branches of physics, it was pointed out that if the structures in $f$ exhibit \textit{scale separation}, in a sense made precise below, they can be encoded both accurately and parsimoniously, on both small and large scales~\cite{Ripoll2021,Ye2022,Gourianov2022,Gourianov2022a,Shinaoka2022}.
This is done using a representation first discussed in the context of quantum information \cite{Wiesner1996,Zalka1998,Grover2002,Latorre2005}, independently introduced in the mathematics literature by Oseledets \cite{Oseledets2009}, and dubbed the 
\textit{quantics} representation by Khoromskij \cite{Khoromskij2011}:
it encodes each 
variable $u_i$ through $R$ binary digits, or bits, and
expresses $f(\bu)$ as a multi-index tensor $f_{\sigma_1 \dots \sigma_\seLL}$, with $\eLL = nR$, where each index represents a bit.
If $f$ exhibits scale separation, this tensor
is highly compressible, i.e.\ it can be well approximated by a tensor train (TT) of fairly low rank. 
These previous works found the TT via singular value decomposition (SVD) of the full tensor, demonstrating
that low-rank quantics TT (QTT) representations \textit{exist}.
It remains to design more practical algorithms to find them, since the computational costs of the SVD approach grow exponentially 
with $\eLL$.
\vspace{-0.18mm}

In an unrelated very recent development
\cite{NunezFernandez2022}, TT representations were used 
for multivariate correlation functions arising in diagrammatic Monte Carlo methods (albeit without using the quantics encoding). It was found that these  TTs are not only highly compressible, but that the compression can be achieved very efficiently using the \textit{tensor cross interpolation} (TCI) algorithm. This technique, 
 pioneered by Oseledets and coworkers \cite{Oseledets2011,Oseledets2010a,Savostyanov2011} and improved by  Dolgov and Savostyanov \cite{Savostyanov2014,Dolgov2020}, is computationally exponentially cheaper than SVDs (albeit theoretically less optimal, though with controlled errors \cite{Savostyanov2014}). 

The purpose of this paper is to point out that 
quantics TTs and TCI can be combined. This leads to a strategy that we call
\textit{quantics tensor cross interpolation} (QTCI). It has several highly desirable properties: 
(i)~Arbitrary resolution via an exponentially large grid with $2^R$ points for each variable, obtained at a cost linear in $R$. (ii)~Efficient construction of the QTCI at cost linear in $\eLL = nR$.
(iii)~Access to many ultrafast algorithms once the QTCI 
has been obtained \cite{Schollwoeck2010,Lubasch2018,Ripoll2021,Shinaoka2022,Garcia-Molina2023}; for instance, 
integrals, convolutions and Fourier transforms can be computed at $\order(\eLL)$ costs (i.e.\ exponentially cheaper than standard fast Fourier transforms)
\cite{supplement}.
\nocite{Dolgov2012,Holzapfel2015,Chen2022,Cortinovis2020,Goreinov2011,Savostyanov2014,NunezFernandez2022,gittoolbox}
(iv)~More generally,
TT representations yield access to a whole range of matrix product states/matrix product operators (MPS/MPO) algorithms which were
devised in the context of many-body physics 
\cite{Schollwoeck2010} and have spawned the mathematical field of TTs.

We illustrate the power of QTCI by using it to resolve 
the momentum dependence of functions defined on the Brillouin zone 
of the celebrated Haldane model \cite{Haldane1988}. We construct QTCIs for its noninteracting Green's function and Berry curvature, and compute the Chern number of a band with topological properties.

\textit{Quantics tensor trains.---}%
For the quantics representation of $f(\bu)$, each variable $u_i$ is 
rescaled to lie within the unit interval $I=[0,1)$, then discretized on a grid of $2^R$ points and expressed as a tuple of \(R\) bits
\cite{Oseledets2009,Khoromskij2011}, \vspace{-1.5mm}
\begin{align}
u_i =  \sum_{b=1}^R \frac{\sigma_{ib}}{2^b} \mapsto (\sigma_{i1} \dots \sigma_{iR}), 
\quad \sigma_{ib} \in \{0,1\} \, . 
\end{align}
\vspace{-2.5mm}

\noindent
Here, $\sigma_{ib}$ resolves the variable $u_i$ at the scale $2^{-b}$.
Arbitrarily high resolution can be achieved by choosing $R$ sufficiently large. 
Thus, $\bu$ is represented by a tuple of $\eLL = n R$ bits.
To facilitate scale separation, the bits are relabled \cite{Oseledets2009} as $\sigma_\ell = \sigma_{ib}$, using a single index $\ell = i + (b-1)n \in 1, \dots, \eLL$. This interleaves them such that all bits $\sigma_{ib}$ 
describing the same scale $2^{-b}$ have contiguous $\sigma_\ell$ labels. Then,  $f$ can be viewed and graphically depicted as tensor of degree $\eLL$:
\begin{flalign}
&	f:  \{0,1\}^\scripteLL \to \mathbbm{C} , 
	\quad 
	\bsigma = (\sigma_1, \dots, \sigma_{\seLL}) \mapsto f_\bsigma =
	f(\bu) \, . \hspace{-1cm} &
\\[1mm] \nonumber
 &    \includegraphics[width=0.933\linewidth]{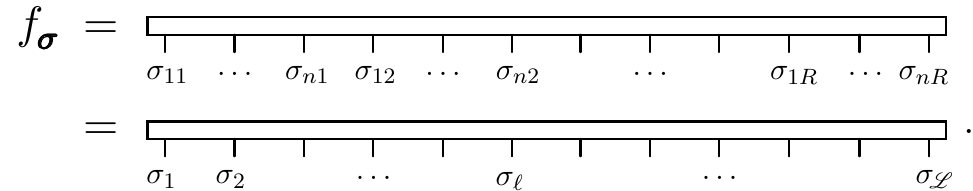}  &
\end{flalign}

\vspace{-2mm} 
\noindent
Alternatively, all same-scale bits can be \textit{fused} together as
\(\tilde{\sigma}_b = \sum_{i=1}^{n} 2^{i-1}  \sigma_{ib} \in \{0, \ldots, 2^n - 1\}\),
yielding the \textit{fused} representation
\(
f_{\tilde{\bsigma}} = f(\bu),
\tilde{\bsigma} = (\tilde{\sigma}_1, \ldots, \tilde{\sigma}_R)
\).
It employs only \(\widetilde \eLL = R\) indices, each of dimension \(d = 2^n\)
\footnote{
    For example, for $n\!=\!2$, $R\!=\!3$, the point
    \((u_1, u_2) = (\tfrac{5}{8},\tfrac{4}{8})\)
    has the binary representation \((101,100)\).
    In the \textit{interleaved} form, the bits are reordered such that $f(\tfrac{5}{8},\tfrac{4}{8})$ is represented by $f_\bsigma = f_{110010}$; in \textit{fused} form, by $f_{\tilde \bsigma} = f_{301}$%
}.

Any tensor can be \textit{unfolded} as a TT \cite{Oseledets2010,Oseledets2010a,Oseledets2011,Khoromskij2011},
graphically depicted as a chain of $\ell$ sites connected by bonds representing sums over repeated indices:
\vspace{-1mm}
\begin{flalign}
\label{subeq:define-MPS}	
f_\bsigma & =  \prod_{\ell=1}^{\seLL}
 M_\ell^{\sigma_\ell}=
[M_1]^{\sigma_1}_{1 \alpha_1} [M_2]^{\sigma_2}_{\alpha_1 \alpha_2} \! \ncdots  
[M_\seLL]^{\sigma_\sseLL}_{\alpha_{\seLL-1} 1}  
\hspace{-1cm} & \\
& =  
     \raisebox{-5mm}{\includegraphics[width=0.833\linewidth]{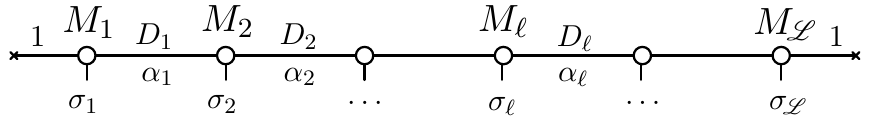}} 
      \vspace{-2cm}  , & 
  \nonumber 
\end{flalign}
\noindent 
Each site $\ell$ hosts a three-leg tensor $M_\ell$ with elements $[M_\ell]^{\sigma_\ell}_{\alpha_{\ell-1} \alpha_\ell}$. 
Its ``local'' and ``virtual'' bond indices, $\sigma_\ell$ and $\alpha_{\ell-1}$, $\alpha_\ell$, have dimensions $d\!=\!2$ and $D_{\ell-1}$, $D_\ell$, respect\-ively,
with $D_0\!=\! D_{\seLL} \!=\! 1$ for the outermost (dummy) bonds. \\
\indent 
If $f_\bsigma$ is full rank, exact TT unfoldings have 
exponential bond growth towards the chain center, $D_\ell = 2^{\min\{\ell, 
\scripteLL -\ell\}}$, implying exponential memory costs, 
$\mathcal{O}(2^{\scripteLL/2})$.
However, tensors $f_\bsigma$ with lower information content admit accurate TT unfoldings with lower virtual bond dimensions. 
Such unfoldings are obtained via iterative factorization and truncation of bonds with low information content. Usually,
this is done using a sequence of SVDs, 
discarding all singular values smaller than 
a specified \textit{truncation threshold} $\epsilon$. 
The largest $D_\ell$ value so obtained, $\Dmax$, is the rank of the $\epsilon$-truncated TT. SVD truncation is provably optimal
\cite{Oseledets2010a}, yielding the smallest possible $\Dmax$ for specified $\epsilon$. If $\Dmax \ll 2^{\scripteLL/2}$, $f_\bsigma$ is \textit{strongly compressible}, implying that it has internal structure. Building on the pioneering studies of Oseledets \cite{Oseledets2010,Oseledets2011} and Khoromskij \cite{Khoromskij2011}, Refs.~\cite{Ye2022,Gourianov2022,Gourianov2022a,Shinaoka2022}
argued that for quantics tensors $f_\bsigma = f(\bu)$, strong compressibility reflects scale separation: structures in $f(\bu)$ occurring on different  scales  are only ``weakly entangled,'' in  that the virtual bonds connecting the corresponding sites in the TT do not require large dimensions. 

This perspective is informed by the study of one-dimensional quantum lattice models using
matrix product states (MPSs)---many-body wave-functions of the form 
 \eqref{subeq:define-MPS} \cite{Schollwoeck2010}. In that context, $\sigma_{\ell}$ labels physical degrees of freedom at site $\ell$, 
and the entanglement of sites $\ell$ and $\ellplusone$ is characterized by an entanglement entropy bounded by $2^{D_\ell}$.
 By analogy, if a quantics TT is strongly compressible, requiring only small $\Dmax$, the sites representing different scales are not strongly entangled---indeed,
 $\Dmax$ quantifies the degree of scale separation inherent in $f(\bu)$.

The SVD unfolding strategy requires knowledge of the full tensor
 $f_\bsigma$: 
 it uses $2^\scripteLL$ function calls, implying exponentially long runtimes, 
 even if $f_\bsigma$ is strongly compressible.
 Thus, this strategy is optimally accurate but exponentially inefficient:
 it uncovers structure in $f_\bsigma$, but 
 does not exploit it already while constructing the unfolded TT.

\textit{Tensor cross interpolation.---}%
The TCI algorithm \cite{Oseledets2010a,Oseledets2011,Dolgov2020,NunezFernandez2022} solves this problem.   
It serves
as a {\it black box} that samples $f_\bsigma$ at some clever choices of \(\bsigma\) and iteratively constructs the TT from the sampled values.
TCI is slightly less accurate than SVD unfoldings, requiring a slightly larger $\Dmax$ for a specified error tolerance $\epsilon$. But it is  
exponentially more efficient, needing at most $\order(\Dmax^2\eLL)$ function evaluations and a run-time of at most \(\order(\Dmax^3\eLL)\).

We refer to \cite{NunezFernandez2022} for details about TCI in general and its actual implementation. Here, we just sketch the main idea.
TCI achieves the factorization needed for unfolding by employing matrix cross interpolation (MCI) rather than SVD. 
Given a matrix $A$, the MCI formula approximates it as \(A \approx C P^{-1} R = \At \), graphically depicted as follows: 
\begin{equation*}
\vspace{-1mm}
    \includegraphics[width=0.98\linewidth]{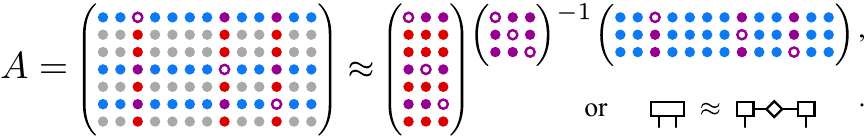}
\vspace{-1mm}
\end{equation*}
Here, the column, row and pivot matrices $C$, $R$, and $P$, are all constructed from elements of $A$: $C$ contains $D$ columns (red), $R$ contains $D$ rows (blue), and $P$ their intersections, the \textit{pivots}  (purple). The resulting $\At$
exactly reproduces all elements of $A$ contained in $C$ and $R$; the remaining elements are in effect interpolated from
the ``crosses'' formed by these (hence ``cross interpolation'').
The accuracy of the interpolation depends on the number and choice of pivots; it can be improved systematically by adding more pivots.
If \(D = \rank(A)\), one can obtain an exact representation of the full matrix,  \(A = \At\) \cite{Oseledets2010a}.

Tensors can be unfolded by iteratively using MCI while treating multiple indices (e.g.\ \(\sigma_2\ldots\sigma_\seLL\)) as a single composite index, e.g.\ 
\(
    f_{\sigma_1\sigma_2\ldots\sigma_{\seLL}} \approx
    [C_1]_{1\beta_1}^{\sigma_1}
    [P_1^{-1}]_{\beta_1\alpha_1}
    [R_1]_{\alpha_1}^{\sigma_2\ldots\sigma_\seLL}
\).
Iterative application of MCI to each new tensor on the right ultimately yields a fully unfolded TT, $f_\bsigma \approx 
f_\bsigma^{\mathrm{QTCI}}$:
\begin{equation*}
    \includegraphics[scale=0.73185]{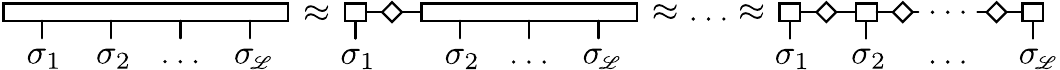} \, \raisebox{4.5mm}{.}
\end{equation*}
[A TT of the form \eqref{subeq:define-MPS} is obtained
by defining $M_\ell = C_\ell P^{-1}_\ell$, 
\raisebox{-1mm}{\includegraphics[scale=1]{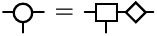}}.] This naive algorithm is inefficient, but illustrates how the interpolation properties of MCI carry over to TCI.
In practice, it is more efficient to use a sweeping algorithm,
successively sampling more function values $f_\bsigma$ and adding pivots to each tensor until the relative error \(\varepsilon\), which decreases during sweeping,
drops below a specified tolerance \(\epsilon\)
\cite{supplement}.
We define $\varepsilon$ as $\max_{\bsigma \in S}|f_\bsigma^{\mathrm{QTCI}} - f_\bsigma|/\max_{\bsigma \in S}|f_\bsigma|$, where $S$ is the set of all $\bsigma$ index values sampled while constructing $f_\bsigma^{\mathrm{QTCI}}$.

\textit{Integration.---}%
The integral over a function \(f\) in QTT form is easily accessible in \(\order(D_{\max}^2 \eLL)\) steps
\cite{NunezFernandez2022,Oseledets2010a}.
It can be approximated by a Riemann sum since the quantics
grid is exponentially fine, and all $\sigma_\ell$ sums 
can be performed independently due to the TT's factorized form:
\begin{align}
   \int_{I^n} \! d^n\vec u f(\vec u) \approx
    \frac{1}{2^{\seLL}} \sum_{\bsigma} f_\bsigma
    \approx \frac{1}{2^{\seLL}}  \prod_{\ell} \sum_{\sigma_\ell} [M_\ell]^{\sigma_\ell}
        \,.
    \label{eq:qttintegral}
\end{align}

\begin{figure}
    \centering
    \includegraphics{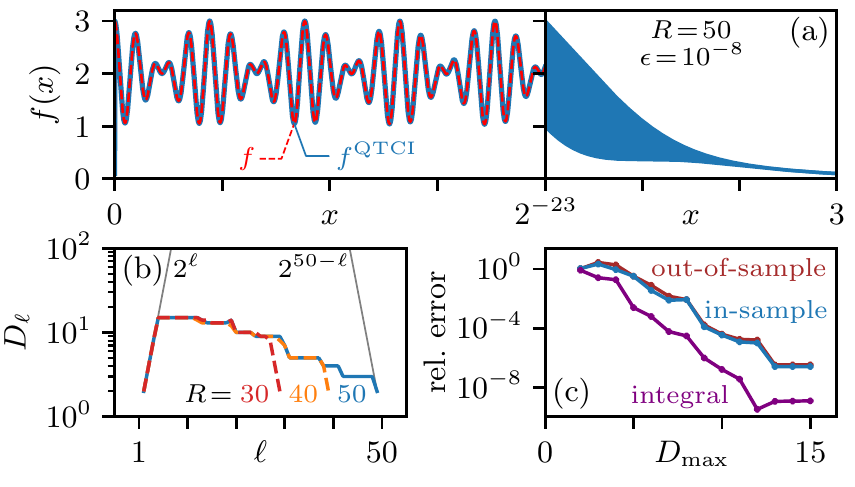}
    \vspace{-7mm}
    \caption{
        QTCI representation of a rapidly oscillating function, for \(\eLL = R = 50\) with tolerance \(\epsilon = 10^{-8}\).
        (a) Plot of \(f(x)\) (see text). Left: the interval \(x \in [0; 2^{-23}]\); red dashed: the actual function, blue: its QTCI representation.
        Right: the envelope structure up to \(x = \log(20)\approx 3\); the rapid oscillations are not resolvable on this scale.
        (b) Virtual bond dimensions \(D_\ell\) of the QTT, for $R=30,40,50$.
        Gray lines indicate how \(D_{\ell}^{R=50}\) would grow without any truncation.
        (c) Relative error estimates as a function of \(D_{\max} = \max (D_\ell)\), for $R=50$. } \vspace{-5mm}
    \label{fig:cosgauss}
\end{figure}

\textit{1D example.---}%
We first demonstrate QTCI for computing 
the integral $I[f]=\int_0^{\ln(20)} dx f(x)$ of the 
function
\(f(x) \!=\!
\cos(\frac{x}{B}) \cos(\frac{x}{4\sqrt{5}B}) \, 
e^{-x^2} \!+ \! 2e^{-x}\) with $B = 2^{-30} \approx 10^{-9}$.
This function, shown in \Fig{fig:cosgauss}(a), involves structure on widely different scales: rapid, incommensurate oscillations and a slowly decaying envelope. A standard representation thereof on an equidistant mesh would  require 
much more than $\order(1/B)$ sampling points, as would the 
computation of the integral $I[f] =
\frac{19}{10} + \order (e^{-1/(4B^2)})$.
By contrast, for a quantics representation,
 it suffices to choose $R$ somewhat larger than $30$ (ensuring $2^{-R} \ll B$);
 and since the information content of $f(x)$ is not very
 high, $f_\bsigma$ is strongly compressible. We unfolded it
 using QTCI with \(\epsilon = 10^{-8}\)  and $R=50$ 
 (quite a bit larger than $30$, just to demonstrate the capabilities of TCI).
 Figure~\ref{fig:cosgauss}(b) shows 
 the resulting profile of $D_\ell$ vs $\ell$, revealing 
 the scale separation inherent in $f(x)$: the initial growth 
 of the bond dimension, $D_\ell \sim e^\ell$, quickly 
 stops at a fairly small maximum, \(D_{\max} = 15\), confirming strong compressibility; thereafter, $D_\ell$ decreases steadily with $\ell$,
 becoming $\mathcal{O}(1)$ for $\ell$ larger than 30, since $f(x)$ has very little structure at scales below $2^{-30}$. 
 Remarkably, although  $f_\bsigma$ has  $2^{50} \approx 10^{15}$ elements, the TCI algorithm finds the relevant structure using 
only \(8706\) samples, i.e.\ roughly \(1\) sample per \(59\,000\) oscillations. 
Nevertheless,
it yields an accurate representation of $f(x)$: 
the in-sample error, the out-of-sample error (defined as maximum error over 2000 random samples)
\footnote{%
    The in-sample error is
    \(\max_{\bsigma\in S} \abs{f^{\text{QTCI}}_\bsigma - f_\bsigma}\),
    where \(S\) is the set of all \(\bsigma\) evaluated during construction of the QTCI. For the out-of-sample error, \(S\) is instead a set of 2000 pseudorandom \(\bsigma\) generated by Xoshiro256++ \cite{Blackman2021}. \emph{Relative} errors are defined as
    \(\max_{\bsigma\in S} \abs{f^{\text{QTCI}}_\bsigma - f_\bsigma} / \abs{f_\bsigma}\)%
},
\nocite{Blackman2021}
and the error for the integral $I[f]$, computed via \Eq{eq:qttintegral}, all decrease exponentially with $\Dmax$
[\Fig{fig:cosgauss}(c)]. The runtimes for computing $I[f]$ using QTCI or adaptive Gauss-Kronrod quadrature are 44~ms vs 6~h on an Intel Xeon W-2245 processor, illustrating the efficiency of QTCI vs conventional approaches.

\textit{Haldane model.---}%
As an example with relevance in physics, we apply QTCI to the Green's function and Berry curvature of the well-known Haldane model \cite{Haldane1988}. It is one of the simplest models with topological properties, yet produces nontrivial structure with multiple peaks and sign changes in reciprocal space. Its Bloch Hamiltonian is 
\begin{multline}\textstyle
    H(\vec k) =
    \sum_{i=1}^3 \Bigl[ \sigma^1 \cos(\vec k \cdot \vec a_i) +
    \sigma^2 \sin(\vec k \cdot \vec a_i) \Bigr] +
    \\+\textstyle
    \sigma^3 \Bigl[
        m - 2t_2 \sum_{i=1}^3 \sin(\vec k \cdot \vec b_i)
    \Bigr],
    \label{eq:haldane_hamiltonian}
\end{multline}
where \(\sigma^\mu\) are Pauli matrices, 
$\vec{k} = (k_x, k_y)$, while \(\vec a_{1,2,3}\) 
connect nearest neighbors and  \(\vec b_{1,2,3}\) next-nearest neighbors of a honeycomb lattice. Compared to Haldane's more general version of the model, we fix his parameters \(\phi = \tfrac{\pi}{2}\), \(t_1 = 1\) and set \(t_2 = 0.1\).
The parameter \(m\) tunes the model through two phase transitions: 
\(\abs{m} < m_\crit = t_2 3\sqrt{3}\) 
yields a Chern insulator with Chern number \(\chern = -1\), and \(\abs{m} > m_\crit \) a trivial phase with \(\chern = 0\) \cite{Haldane1988}.
At  $m = \pm m_\crit$, a single Dirac point appears at \(\vec k = (\mp \frac{4}{3}\pi, 0)\) and symmetry-related \(\vec k\);  there, the Chern number is \(\chern = -\tfrac{1}{2}\) \cite{Watanabe2011}.

\begin{figure}
    \centering
    \includegraphics{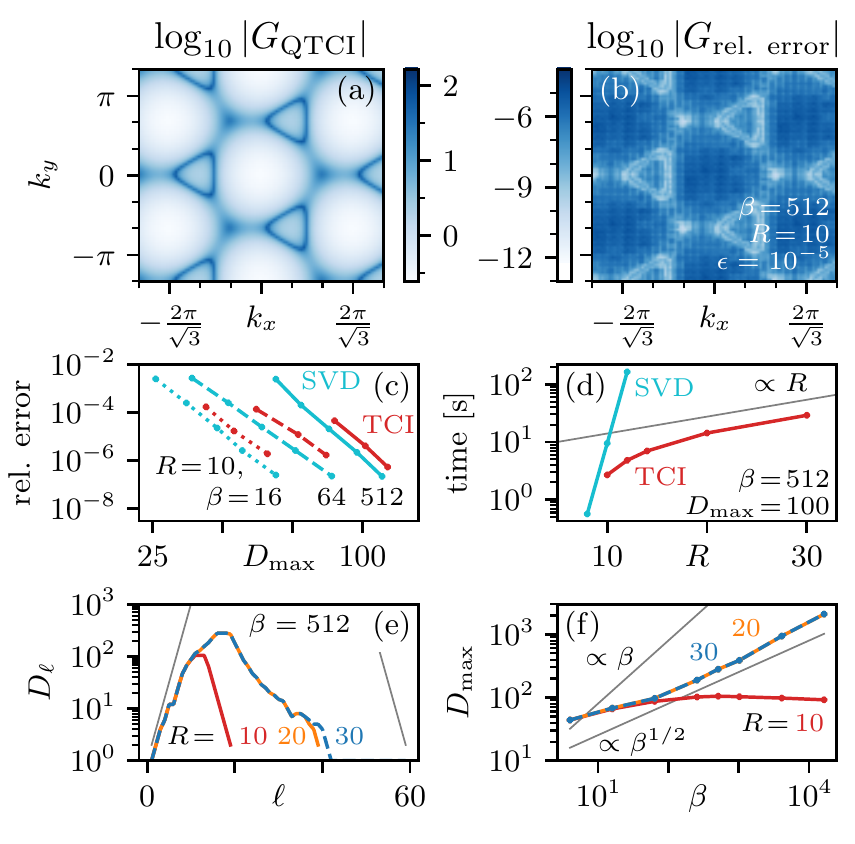}
    \vspace{-7.5mm}
    \caption{
    Green's function $G(\bk)$ of the Haldane model, computed with error tolerance \(\epsilon = 10^{-5}\) throughout. (a,b)~QTCI of the $|G|$ and its relative error, \(|G_\mathrm{QTCI} - G| / |G|\), for \(\beta = 512\), \(R = 10\). 
    (c,d)~Comparison of QTT unfoldings via SVD and TCI, showing  (c) the relative error vs the maximum bond dimension for $R=10$ and  \(\beta=16, 64, 512\); and (d) runtimes vs $R$ for $\beta = 512$. 
    (e,f) QTCI bond dimensions for \(R = 10, 20, 30\), showing (e) $D_\ell$ vs $\ell$ for \(\beta = 512\); and (f) \(D_{\max}\) vs \(\beta\).}
\vspace{-4mm}
    \label{fig:dyson}
\end{figure}

\textit{Green's function in reciprocal space.---}%
To illustrate QTCI for the Haldane model, we study the momentum dependence of the Green's function,
$G(\vec{k}, i\omega_0) = \Tr[(i\omega_0 - H(\vec{k}) + \mu)^{-1}]$,
with \(\omega_0 = \pi/\beta\) the lowest fermionic
Matsu\-bara frequency and \(\Tr\) 
traces over the \(2\!\times\!2\) space of $H$.

 Figure~\ref{fig:dyson}(a) shows an intensity plot of the QTCI representation of \(G\) in reciprocal space; \Fig{fig:dyson}(b) shows 
 that the relative error 
with respect to the exact value is below $10^{-5}$ throughout,
hence the momentum dependence is captured accurately.
There are small Fermi surfaces around
\(\vec k = (-\tfrac{4}{3}\pi, 0)\) and symmetry-related \(\vec k\).
To construct QTTs, we define $f_{\bsigma} = G(\vec{k}, i\pi/\beta)$, where 
\(\bsigma\) encodes \(\bk\)
and \(\beta\) is fixed. Figure~\ref{fig:dyson}(c) shows the relative in-sample error as a function of \(\Dmax\) for TTs constructed with \(R = 10\) for $\beta=16$, $64$, $512$, using either SVD or TCI. For both, the error decreases exponentially as $\Dmax$ increases. 
Moreover, TCI is nearly optimal, achieving the same error
as SVD for a $\Dmax$ that is only a few percent larger.

Figure~\ref{fig:dyson}(d) shows how SVD and TCI 
runtimes depend on the number of bits, $R$, for a fixed $\Dmax$ at 
large $\beta=512$, where the features in $G$ are sharp.
The times, including function evaluations, were measured on a single CPU core of AMD EPYC 7702P.
The SVD runtimes become prohibitively large for $R>10$ due to exponential scaling; by contrast, the TCI runtimes depend only mildly on \(R\).

Figure~\ref{fig:dyson}(e) shows how TCI profiles of \(D_{\ell}\) vs $\ell$ depend on \(R\), for \(\beta=512\) and a specified error tolerance \(\epsilon = 10^{-5}\). The bond dimension
initially grows as \(D_{\ell} \sim 2^\ell\), reaches a maximum near \(\ell \approx 20\), then decreases back to \(1\).
The curves for \(R \!=\! 20\) and \(30\) almost coincide, indicating that a good resolution of the sharp features at \(\beta = 512\) requires  \(R > 20\)---well beyond the reach of SVD unfoldings.

The low computational cost of TCI allows us to investigate the $\beta$ dependence of $\Dmax$, easily reaching \(\beta\!=\!2^{14}\!=\!16384\).
Figure~\ref{fig:dyson}(f) suggests $\Dmax \propto \beta^\alpha$ with $\alpha \approx 1/2$ for large $\beta$. 
Remarkably, this growth is slower than that, $\Dmax \propto \beta$, conjectured  for a scheme based on SVD and patching \cite{Shinaoka2022}. A detailed analysis for general models and higher spatial dimensions is an interesting topic for future research. 

\textit{Chern number.---}%
Finally, we consider the Chern number, \(\chern\), for the Haldane model at \(\mu = 0\) and $\beta = \infty$.
To avoid cumbersome gauge-fixing procedures, we use the gauge-invariant method described in Ref.~\cite{Fukui2005}.
First, we discretize the Brillouin zone (BZ)
into
\(2^R \!\times \! 2^R\) plaquettes. 
Then,
the Chern number can be obtained from a sum
over plaquettes,
\(
    \chern \approx \frac{1}{2\pi i} \sum_{\vec k \in \text{BZ}} \berry(\vec k)
\),
where 
\(
\berry(\vec k) \approx
-i\arg(
\braket{\psi_{\vec k_1}}{\psi_{\vec k_2}}
\braket{\psi_{\vec k_2}}{\psi_{\vec k_3}}
\braket{\psi_{\vec k_3}}{\psi_{\vec k_4}}
\braket{\psi_{\vec k_4}}{\psi_{\vec k_1}}
)
\) is the Berry flux through the plaquette with corners 
\(\vec k_1\ldots \vec k_4\), and \(\ket{\psi_{\vec k}}\) are valence band wave functions.

\begin{figure}
    \centering
    \includegraphics{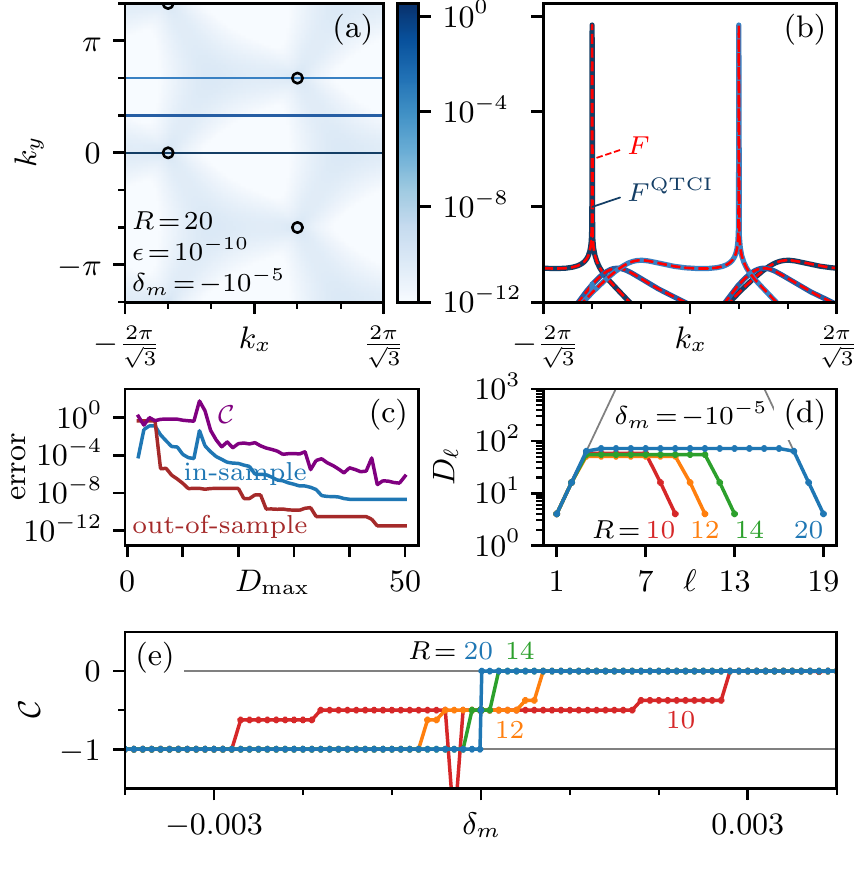}
    \vspace{-8.5mm}
    \caption{
    Evaluation of the Chern number in the Haldane model using QTCI for the Berry flux \(\berry(\vec k)\), with error tolerance $\epsilon = 10^{-10}$ throughout.
    (a)~QTCI of \(\berry(\vec k)\) on the integration domain (circles mark peak positions);
    (b)~cuts through \(\berry(\vec k)\) along the colored lines shown in (a);
     and  (c)~errors for $\berry(\vec k)$ and 
    \(\chern\) as functions of \(D_{\max}\), all computed for \(\deltam \!=\! -10^{-5}\), 
    \(R \!=\! 20\).
    (d)~Bond dimension \(D_\ell\) for QTTs of length \(R = 10, 12, 14, 20\). 
    (e)~Chern number \(\chern\) as a function of \(\deltam\), for four choices of $R$.}
    \vspace{-5mm}
    \label{fig:haldane}
\end{figure}

Close to the transition, for small \(\deltam = m - m_\crit\), the band gap is \(2\deltam\). This induces peaks of width \(\sim\deltam\) in the Berry flux \(\berry(\vec k)\), shown in \Figs{fig:haldane}(a, b) for \(\deltam = 10^{-5}\). There, we used a fused
quantics representation with $R=20$,
ensuring a mesh spacing $2^{-R}$ well smaller than $\delta_m$.
Whereas a calculation of $\chern$ via direct summation or SVD unfolding would require $2^{2R} \approx 10^{12}$ function evaluations, QTCI is much more efficient: for 
a relative tolerance of \(\epsilon = 10^{-10}\), it needed only \(4\times 10^5\) samples (and  20~s runtime on a single core of an Apple M1 processor). It yielded a QTT with maximum bond dimension \(D_{\max} = 50\), and a Chern number within $10^{-6}$ of the expected value \(\chern = -1\) 
(see Figs.~\ref{fig:haldane}(c,d)). 
When plotted as a function of $\delta_m$,
$\chern$ shows a sharp step from $-1$ to $0$ at \(\deltam = 0\) if computed using $R=20$ (\Fig{fig:haldane}(e)), beautifully demonstrating that the $\bk$ mesh is fine enough. For smaller $R$ the mesh becomes too coarse, incorrectly yielding a plateau at $-1/2$ instead of a sharp step.

For benchmarking purposes, we deliberately chose a model that is analytically solvable. 
However, our prior knowledge of the peak positions of the Berry curvature
was not made available to TCI. This demonstrates its reliability in finding sharp structures, provided enough quantics bits are provided to resolve them.
Random sampling misses these sharp structures, which is why in Fig.~\ref{fig:haldane}(c) the out-of-sample error, obtained from 2000 random samples, lies well below the in-sample error.

\textit{Outlook.---}%
We have shown that the combination of the quantics representation~\cite{Wiesner1996,Zalka1998,Grover2002,Latorre2005,Ripoll2021,Ye2022,Gourianov2022,Gourianov2022a,Oseledets2009,Khoromskij2011,Shinaoka2022} with TCI \cite{Oseledets2010a,Oseledets2011,Dolgov2020,NunezFernandez2022} is a powerful tool for
uncovering low-rank structures in exponentially large, yet very common objects: functions of few variables resolved with high resolution. 
Numerical work with such objects always involves truncations---the radically new perspective opened up by QTCI is that they can be performed at \textit{polynomial costs} by discarding weak entanglement between different scales. 
Once a low-rank QTT has been found, it may be further used within one of the many existing MPO/MPS algorithms \cite{Schollwoeck2010,Lubasch2018,Ripoll2021,Shinaoka2022,Garcia-Molina2023}.

We anticipate that the class of problems for which QTCI can be instrumental is actually very large,  reaching well beyond the scope of physics. Intuitively speaking, the only requirement is that the functions should entail some degree of scale separation and not be too irregular (since random structures are not compressible). Thus, a large new research arena, potentially connecting numerous different branches of science, awaits exploration. Fruitful challenges: 
establish criteria for which types of multivariate functions admit low-rank QTT representations; develop improved algorithms for constructing low-rank approximations to tensors; and above all, explore the use of QTCI 
for any of the innumerable problems in science requiring high-resolution numerics. The initial diagnosis is easy: simply use SVDs or QTCI \cite{gittoolbox} to check whether the
functions of interest are compressible or not.

\begin{acknowledgements}
We thank Takashi Koretsune, Ivan Oseledets and Bj\"orn Sbierski for inspiring discussions, and Jeongmin Shim for important help at the beginning of this work. We carried out part of the calculations using computer code based on ITensors.jl~\cite{Fishman2022} written in Julia~\cite{Bezanson2017}.
H.\,S. was supported by JSPS KAKENHI Grants No. 21H01041, and No. 21H01003, and JST PRESTO Grant No. JPMJPR2012, Japan. X.\,W. acknowledges funding from the  Plan France 2030 ANR-22-PETQ-0007 ``EPIQ''; and J.\,v.\,D. from the Deutsche Forschungsgemeinschaft under Germany's Excellence Strategy EXC-2111 (Project No. 390814868), and the Munich Quantum
Valley, supported by the Bavarian state government
with funds from the Hightech Agenda Bayern Plus.
\end{acknowledgements}

\vspace{-4mm}
\bibliography{QTCI}

\begin{thebibliography}{36}
\bibitem [{git()}]{gittoolbox}%
  \BibitemOpen
  \href@noop {} {\bibinfo  {journal} {A ready-for-use QTCI toolbox will be
  published as open source library in the near future}\ }\BibitemShut {NoStop}%
\end{thebibliography}
%


\supplement

This supplement offers more detail on the runtime complexity of calculating a Fourier transform using QTTs in Sec.~\ref{sec:supp:FT}, and a brief review of the TCI algorithm in Sec.~\ref{sec:supp:TCI}.

\section{Fourier transform of a QTT}
\label{sec:supp:FT}

This section explains how to compute the Fourier transform of a function in QTT form to substantiate the claim that this can be done faster than fast Fourier transform (FFT). The algorithm is related to the so-called superfast Fourier transform \cite{Dolgov2012} and quantum Fourier transform \cite{Holzapfel2015}, and has been formulated in matrix product operator (MPO) form in Refs.~\cite{Chen2022,Shinaoka2022}.

In quantics representation, the Fourier transform can be approximated as matrix product operator (MPO) with very small bond dimension,
which can then be applied to a function in QTT form at a cost of \(\order(D^3\eLL)\) including recompression \cite{Shinaoka2022,Holzapfel2015,Chen2022}.
The cost of constructing the MPO is in most cases much smaller then the cost of applying it to the QTT.
For QTT with moderate bond dimension \(D\), the QTT Fourier transform can be orders of magnitude faster than FFT \cite{Dolgov2012,Chen2022}.
For a generic non-compressible function, by contrast,  the bond dimension will grow exponentially, which results in the same worst-case complexity as using FFT (\(O(N\log N)\) with \(N = 2^\seLL\)).
A detailed analysis including numerical experiments can be found in Ref.~\cite{Chen2022}.

\section{Tensor Cross Interpolation}
\label{sec:supp:TCI}

This section gives an overview of the sweeping algorithm we used to construct a tensor cross interpolation (TCI) in practice. For details, see Ref.~\cite{NunezFernandez2022}.

\begin{figure}[b]
    \centering
    \includegraphics[scale=0.73185]{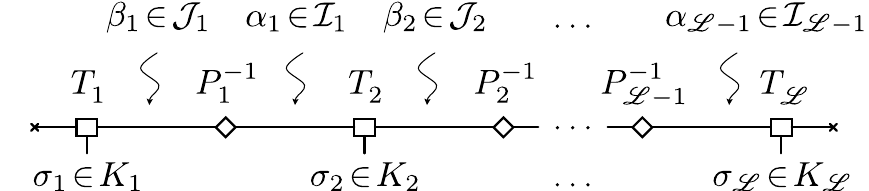}
    \caption{Structure of the TCI during optimization in graphical notation. Squares denote 3-leg tensors \(T_\ell\), diamonds denote matrices \(P_\ell^{-1}\). The indices \(\sigma_\ell\) are external, while the indices \(\alpha_\ell\) and \(\beta_\ell\) are shared between neighbouring \(T_\ell\) and \(P_\ell\).}
    \label{fig:supp:TCIstructure}
\end{figure}

Given a tensor \(f_\bsigma\) with indices~\(\bsigma = (\sigma_1, \ldots, \sigma_\seLL)\) and \(\sigma_\ell \in K_\ell := \{0, \ldots, d_\ell-1\}\), our goal is to construct a TCI approximation which fulfills a user-specified error tolerance~\(\epsilon\) with high efficiency.
The TCI is assembled from three-leg tensors~\([T_\ell]^{\sigma_\ell}_{\alpha_{\ell-1}\beta_{\ell}}\)
and two-leg matrices~\([P_\ell^{-1}]_{\beta_{\ell}\alpha_{\ell}}\) as
\begin{align}
&
f^{\text{QTCI}}_{\sigma_1\ldots\sigma_{\seLL}} =
\label{eq:supp:QTCIconstruction}
\\\qquad
[T_1]^{\sigma_1}_{\alpha_0\beta_1}
[P_1^{-1}]_{\beta_1\alpha_1}
[T_2]^{\sigma_2}_{\alpha_1\beta_2}
[P_2^{-1}]_{\beta_2\alpha_2}
\ldots
[T_\seLL]^{\sigma_\seLL}_{\alpha_{\seLL-1}\beta_\seLL}
\,,
\span
\notag
\end{align}
where \(\alpha_0 = 1\) and \(\beta_\seLL = 1\) are dummy indices, and repeated indices are summed over (see Fig.~\ref{fig:supp:TCIstructure}).
Using \emph{multi-index sets}
\(\mathcal{I}_\ell \subseteq K_1 \otimes \ldots \otimes K_\ell\)
and
\(\mathcal{J}_\ell \subseteq K_\ell \otimes \ldots \otimes K_\seLL\),
all components of the TCI can be described as images of these multiindex sets:
\begin{equation}
    T_\ell =
    f_{\mathcal{I}_{\ell-1} \otimes K_\ell \otimes \mathcal{J}_{\ell}}
    ;\qquad
    P_\ell^{-1} =
    (f_{\mathcal{I}_{\ell} \otimes \mathcal{J}_{\ell}})^{-1}
    \label{eq:supp:TPdefinition}
\end{equation}
Then, the indices \(\alpha_\ell\) enumerate elements of \(\mathcal{I}_\ell\) in Eq.~\eqref{eq:supp:QTCIconstruction}, and \(\beta_\ell\) enumerate elements of \(\mathcal{J}_\ell\). The problem of finding a good TCI, i.e.~good \(T_\ell\) and \(P_\ell\), has now been reduced to optimizing \(\mathcal{I}_\ell\) and \(\mathcal{J}_\ell\).
The cost of the algorithm is determined by the \emph{bond dimensions} \(D_\ell = \abs{\mathcal{I}_\ell} = \abs{\mathcal{J}_\ell}\), which should optimally be kept as small as possible, while still large enough to approximate \(f_\bsigma\) well.

We initialize the TCI starting from an arbitrary initial pivot point \((\initsigma_1, \ldots, \initsigma_\seLL)\) typically provided by the user.
Now, set
\(\mathcal{I}_\ell = \{(\initsigma_1, \ldots, \initsigma_\ell)\}\)
and
\(\mathcal{J}_\ell = \{(\initsigma_\ell, \ldots, \initsigma_\seLL)\}\).
Each \(T_\ell\) now contains the full dependence of \(f_\bsigma\) on a specific index \(\sigma_\ell\), but only takes a single value \(\initsigma_{\ell'}\) of all other digits \(\sigma_{\ell'\neq\ell}\) into account. This corresponds to exactly one \emph{pivot} in MCI language.
These \(T_\ell\) and \(P_\ell\) are now sucessively refined by adding pivots as follows.

The algorithm sweeps over the TCI repeatedly.
During each sweep,
two-site optimizations are applied to bonds between neighbouring tensors \(T_\ell\) and \(T_{\ell+1}\), where
\(\ell\) progresses from \(1\) to \(\eLL-1\) for forward sweeps, and vice versa for backward sweeps.
For local optimization of bond \(\ell\), the algorithm first evaluates a two-site tensor
\([\Pi_\ell]_{\alpha \sigma_\ell; \sigma_{\ell+1} \beta} = f_{\alpha \sigma_{\ell} \sigma_{\ell+1} \beta}\),
where \(\alpha \in \mathcal{I}_{\ell-1}\), \(\sigma_{\ell} \in K_{\ell}\), \(\sigma_{\ell+1} \in K_{\ell+1}\) and \(\beta \in J_{\ell+2}\).
It describes the dependence of \(f_\bsigma\) on the indices~\(\sigma_\ell\) and \(\sigma_{\ell+1}\) exactly, but considers only one value of the other indices, namely the value specified by \(\alpha\in\mathcal{I}_{\ell-1}\) and \(\beta\in\mathcal{J}_{\ell+2}\).
We fuse the index \(\alpha\) with \(\sigma_\ell\) and index \(\sigma_{\ell+1}\) with \(\beta\), such that \(\Pi_\ell\) is reshaped into a matrix. Then, we perform an MCI
\begin{equation}
    [\Pi_\ell]_{\alpha_{\ell-1} \sigma_\ell; \sigma_{\ell+1} \beta_{\ell+2}}
    \approx
    C_{\alpha_{\ell-1} \sigma_{\ell}; j}
    \,
    [P^{-1}]_{ji}
    \,
    R_{i; \sigma_{\ell+1} \beta_{\ell+1}}
    \,,
    \label{eq:supp:PiMCI}
\end{equation}
where \(C\) contains \(D_\ell\) columns of \(\Pi_\ell\), \(R\) contains \(D_\ell\) rows, and \(P\) their intersection points:
\begin{equation*}
    \includegraphics[width=0.98\linewidth]{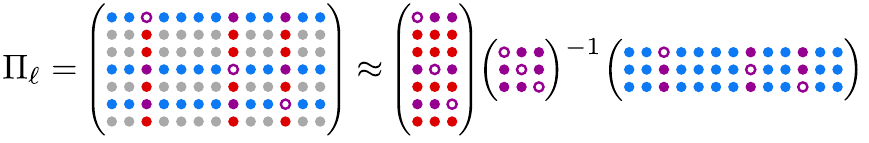}
\end{equation*}
For a given bond dimension \(D_\ell\), the error is quasi-optimal if \(P\) is chosen such that \(\abs{\det{P}}\) is maximal (maximum volume principle) \cite{Goreinov2011,Cortinovis2020}.
We construct \(C, P\) and \(R\) iteratively, where in each iteration we add the row and column that maximize \(\abs{\det{P}}\). The iteration continues until the difference between the left- and right-hand side of Eq.~\eqref{eq:supp:PiMCI}, \(\varepsilon_\ell = \norm{\Pi_\ell - CP^{-1} R}_{\infty}\), is smaller than a user-specified tolerance~\(\epsilon_{\text{MCI}}\). This is equivalent to a fully pivoted rank-revealing LU decomposition, which is used to construct the MCI in practice~\cite{Cortinovis2020}.

Since \(\Pi_\ell\) is a subset of the \(f_\sigma\) tensor, the choice of rows and columns in the MCI corresponds to choosing a subset of important multi-indices at bond \(\ell\). Therefore, we update \(\mathcal{I}_{\ell}\) by including the set of rows that form \(R\),
and \(\mathcal{J}_{\ell}\) by including the set of columns that form \(C\), possibly increasing the number of elements in those sets. Concretely, the tensors \(T_\ell, T_{\ell+1}\) are updated as
\begin{align*}
    [T_\ell]^{\sigma_\ell}_{\alpha_{\ell-1}\beta_{\ell}}
    &\leftarrow
    C_{\alpha_{\ell-1} \sigma_\ell; \beta_{\ell}}
    \,,&
    [P_\ell^{-1}]_{\beta_{\ell} \alpha_{\ell}}
    &\leftarrow
    [P^{-1}]_{\beta_{\ell} \alpha_{\ell}}
    \,,\\
    [T_{\ell+1}]^{\sigma_{\ell+1}}_{\alpha_{\ell}\beta_{\ell+1}}
    &\leftarrow 
    R_{\alpha_{\ell}; \sigma_{\ell+1} \beta_{\ell+1}}
    \,.
\end{align*}
This preserves Eq.~\eqref{eq:supp:TPdefinition}, now with updated \(\mathcal{I}_\ell\) and \(\mathcal{J}_\ell\).

Local optimizations are performed successively for each bond during the sweep, such that one full sweep corresponds to optimization of all \(\mathcal{I}_\ell\) and \(\mathcal{J}_\ell\). The algorithm continues sweeping back and forth until the bond dimension
\(\Dmax = \max_\ell D_\ell = \max_\ell \abs{\mathcal{I}_\ell}\) is converged, and a global error estimate \(\varepsilon \approx \max_\bsigma \abs{f^{\text{QTCI}}_\bsigma - f_\bsigma}\) is smaller than a user-specified tolerance \(\epsilon\).
The global error estimate is given by \(\varepsilon = \max_\ell \varepsilon_\ell\),
where \(\varepsilon_\ell\) is the maximum absolute difference between the left- and right-hand-side of Eq.~\eqref{eq:supp:PiMCI}.
These local errors~\(\varepsilon_\ell\) are a good estimate of the global error because the multi-indices in \(\mathcal{I}_\ell\) and \(\mathcal{J}_\ell\) are \emph{nested}, i.e.~
\(\mathcal{I}_\ell \subseteq \mathcal{I}_{\ell-1} \otimes K_\ell\)
and
\(\mathcal{J}_\ell \subseteq K_\ell \otimes \mathcal{J}_{\ell+1}\)
\cite{NunezFernandez2022,Savostyanov2014}.

In practice, it is advantageous to choose the MCI tolerance~\(\epsilon_{\text{MCI}}\) slightly smaller than the TCI tolerance~\(\epsilon\), and the initial pivot~\(\initsigma_\ell\) close to the largest structures in \(f_\bsigma\). An optimized library that was used for the examples in the main text will be released in the near future \cite{gittoolbox}.

\end{document}